# Treatment Planning System for Electron FLASH Radiotherapy: Open-source for Clinical Implementation

Mahbubur Rahman[1*], M. Ramish Ashraf[1], David J. Gladstone[1,2,3], Petr Bruza[1], Lesley A. Jarvis[2,3], Philip E. Schaner[2,3], Xu Cao[1], Brian W. Pogue[1,3,4], P. Jack Hoopes[1,2,3,4], Rongxiao Zhang[1,2,3*]

[1] Thayer School of Engineering, Dartmouth College, Hanover NH 03755, US

[2] Department of Medicine, Radiation Oncology, Geisel School of Medicine, Dartmouth College Hanover NH 03755 USA

[3] Norris Cotton Cancer Center, Dartmouth-Hitchcock Medical Center, Lebanon, NH 03756 USA

[4] Department of Surgery, Geisel School of Medicine, Dartmouth College, Hanover NH 03755 USA

[*]Correspondence to: Mahbubur.Rahman.TH@dartmouth.edu or Rongxiao.Zhang@dartmouth.edu

**Abstract**

**Purpose:** A Monte Carlo (MC) beam model and its implementation in a clinical treatment planning system (TPS, Varian Eclipse) are presented for a modified ultra-high dose-rate electron FLASH radiotherapy (eFLASH-RT) LINAC.

**Methods:** The gantry head without scattering foils or targets, representative of the LINAC modifications, was modelled in Geant4. The energy spectrum ($\sigma_E$) and beam source emittance cone angle ($\theta_{cone}$) were varied to match the calculated and Gafchromic film measured central-axis percent depth dose (PDD) and lateral profiles. Its Eclipse configuration was validated with measured profiles of the open field and nominal fields for clinical applicators. eFLASH-RT plans were MC forward calculated in Geant4 for a mouse brain treatment and compared to a

Clinical Treatment Planning System for eFLASH RT
conventional (Conv-RT) plan in Eclipse for a human patient with metastatic renal cell carcinoma.

**Results:** The beam model and its Eclipse configuration agreed best with measurements at $\sigma_E$=0.5 MeV and $\theta_{cone}$ =3.9±0.2 degrees to clinically acceptable accuracy (the absolute average error was within 1.5% for in-water lateral, 3% for in-air lateral, and 2% for PDD). The forward dose calculation showed dose was delivered to the entire mouse brain with adequate conformality. The human patient case demonstrated the planning capability with routine accessories in relatively complex geometry to achieve an acceptable plan (90% of the tumor volume receiving 95% and 90% of the prescribed dose for eFLASH and Conv-RT, respectively).

**Conclusion:** To the best of our knowledge, this is the first functional beam model commissioned in a clinical TPS for eFLASH-RT, enabling planning and evaluation with minimal deviation from Conv-RT workflow. It facilitates the clinical translation as eFLASH-RT and Conv-RT plan quality were comparable for a human patient. The methods can be expanded to model other eFLASH irradiators.

## 1. Introduction

There is increasing evidence that ultra-high dose rate (UHDR, >40 Gy/s) treatment delivery to patients can lead to the FLASH effect[1–3], or improved therapeutic ratio by reducing normal tissue toxicity[4–7]. While the first human patient was treated with an electron FLASH beam at UHDR and several preventative dosimetry checks were done to ensure safe delivery[2], widespread translation to the clinic would benefit from prior prediction of dose to patients, which necessitates a treatment planning process to account for anatomical heterogeneity and complex geometries, predict tumor and organs-at-risk dose and create deliverable plans in a clinical setting[8].

In recent FLASH RT studies, Van de Water et al. utilized their in-house treatment planning system to investigate their proton pencil beam scanning system's potential for FLASH RT delivery[9]. Van Marten et al. evaluated efficacy of FLASH RT treatments for a passive proton beam on the widely used clinical Eclipse TPS (Varian Medical Systems, Palo Alto, CA). In a recent study by Rahman et al.[10] UHDR megavoltage electron beams were delivered to isocenter on a modified medical linac in its normal clinical setting. The described modifications enabled the delivery of UHDR beams with minimal modifications to the machine configuration. A similarly modified TPS is desirable to model characteristics of the UHDR beam that are different from conventional beams.

Monte Carlo (MC) dose calculation can reduce the dose prediction uncertainty to about 2-3% even in complex treatment geometries and patient compositions[11]. MC methods have been widely adopted in modern treatment planning systems (TPS) for accurate dose calculation and plan optimization. In this study, a Monte Carlo (MC) model of an UHDR electron beam was developed on the Geant4-based GAMOS MC toolkit and implemented into the Eclipse TPS for planning treatments to biological subjects (process shown in Figure 1). The model was representative of a clinical linear accelerator (Varian Clinac 2100 C/D) modified to deliver UHDR at treatment room isocenter utilizing clinical accessories and geometry[10]. The beam's parameters (e.g. mean energy, energy spread, spot size or spatial spread, and cone angle) were determined to best match film measured dose profiles. The measured profiles were used to configure the Eclipse eMC TPS beam model and both models were validated with film measurements. As demonstrations, the predicted dose distributions in a human metastatic renal cell carcinoma and mouse whole brain treatments and potential clinical importance in adoption of electron FLASH-RT (eFLASH-RT) treatment planning were presented.

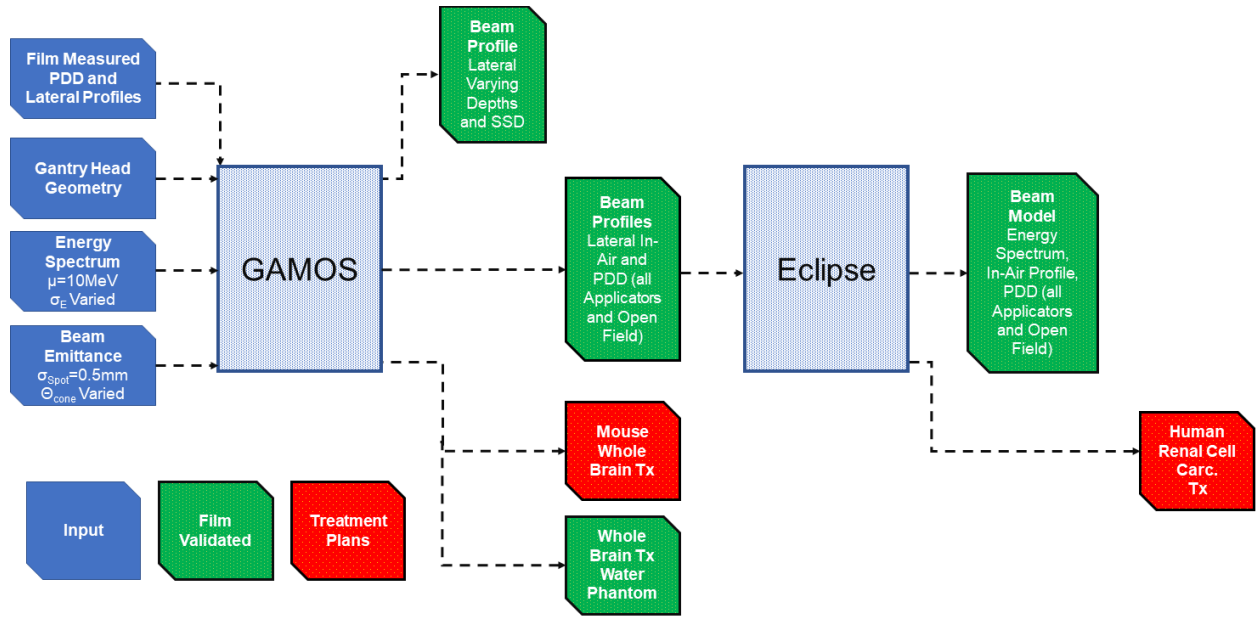

**Figure 1.** Methods to develop and validate beam models and create treatment plans on GAMOS MC toolkit and Eclipse TPS. The diagram can be read as a timeline from left to right.

## 2. Material and Methods

A diagram of methods used to develop the beam model and treatment plans of the biological subjects are included in Figure 1. eFLASH beam from the modified Varian Clinac 2100 C/D at UHDR (ultra-high dose rates of ~300 Gy/s, or ~1 Gy/pulse) irradiated EBT XD Gafchromic film (Ashland Advanced Materials, Bridgewater, NJ) to characterize the beam's spatial dose distribution. The LINAC delivered 40 pulses per acquisition. The films were placed orthogonal to beam axis, on top of or in between varying thickness of solid water phantom with a slab of 5cm solid water phantom (Sun Nuclear, Florida, USA) below for backscatter. The percent depth dose (PDD) at discrete depths were determined with 2.5×2.5 cm$^2$ film placed along the central axis. The lateral dose profiles were measured using 25×20 cm$^2$ film placed at the surface, 1cm, 3cm and 4 cm depths at 100 cm source to surface distance (SSD). The surface lateral profiles

Clinical Treatment Planning System for eFLASH RT

were also measured at 95 cm, 110cm, and 120 cm SSD. The films were read out and dose was quantified using the methods described in Rahman et al.[10] (2021).

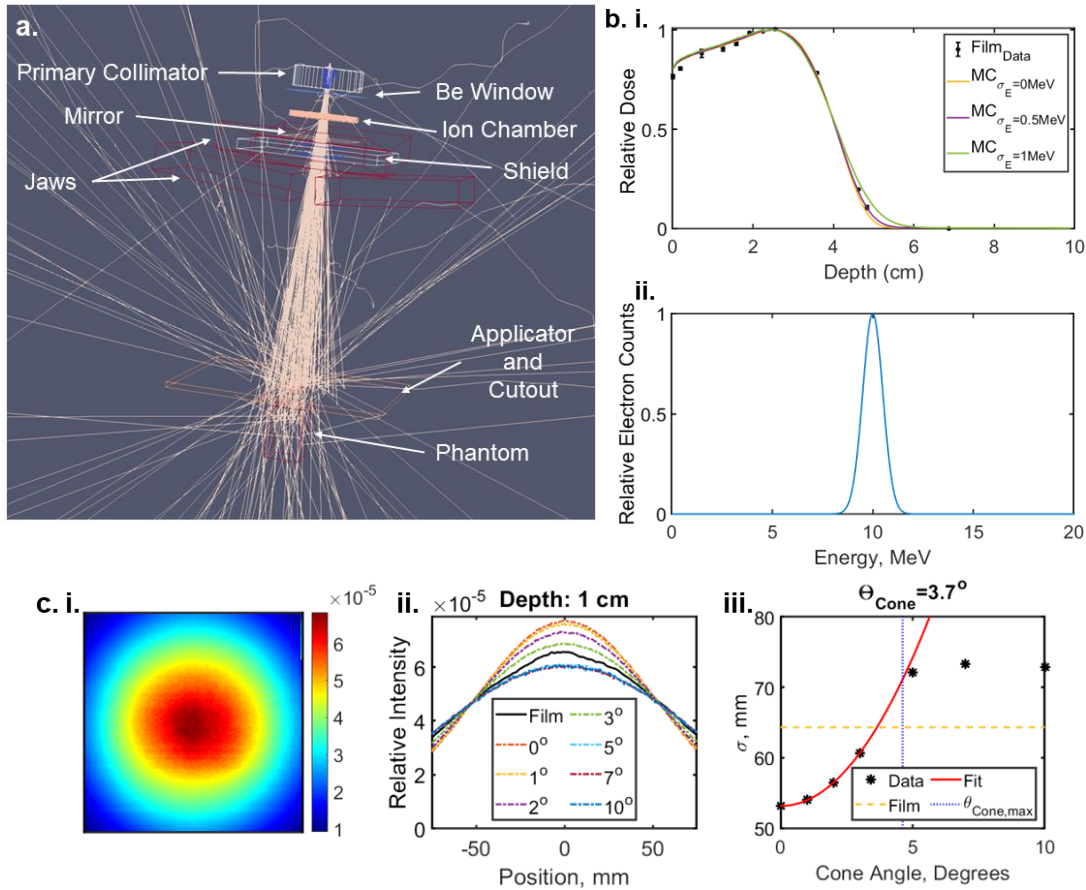

**Figure 2. a.** Set up for MC simulations of the electron FLASH beam. Methods of determining the beam parameters **b.** energy spectrum standard deviation and **c.** emittance cone angle with an example at 1cm depth and in-plane lateral profile. The implemented energy spectrum is included in **b.ii.** The average of cone angles determined from the fits and the film at different depths for the lateral profiles was chosen as the eFLASH beam's cone angle (each calculated cone angle and plots are included in Supplementary Figure 1A).

The emission parameters that best represented the measured film dose profiles were determined by modelling the head of the LINAC on the GEANT4 based GAMOS MC (6.1.0) toolkit. Figure 2a shows the model of the LINAC head with the scattering foil, flattening filter, and the target removed, representative of the modified clinical LINAC used to deliver UDHR beams[10] in a previous study. For the unperturbed 10 MeV electron beam produced by the LINAC, the standard deviation ($\sigma_E$) of the nominal beam energy was varied to determine which best fit the PDD measured by the film as shown in figure 2b. The methods for determining the beam emittance using the open field beam (jaws wide open with a 40×40 cm$^2$ field size setting) are shown in figure 2c (the energy spectrum that best matched in figure 2b was implemented). Prior studies have shown the 0.5 mm (1.18 mm FWHM)[12] spot size, $\sigma$ or standard deviation in profile at the source, for a gaussian emittance distribution of the electron beam is best representative of the LINAC. The $\sigma$ of lateral dose profiles for the beam at 100 cm SSD were compared to that from the MC results at varying cone angles. The quadratic fit was applied to determine the cone angle that would equate the $\sigma$ from the MC results to that measured[13] from the film. (Note: the fit did not include simulated cone angles above 4.63 degrees and discussed further in the results). The appropriate cone angles were determined for each measured depth along the in-plane and cross-plane and averaged as shown in supplementary figure 1A. Enough particles were simulated to quantify the dose distribution with <1% uncertainty (from the entrance to the practical range of the electrons) in the open field beam for each beam parameter variation with a 25×25×10 cm$^3$ water phantom and 1×1×1 mm$^3$ voxel size. The lateral dose profile at each measured depth and SSD were compared with the MC results (from simulating with the determined cone angle and standard deviation in the energy spectrum) as shown in figure 3.

With confirmed beam parameters from film measured lateral profiles, an Eclipse eMC TPS beam model was created from the validated GAMOS model. Open field and the applicators (i.e. 6×6cm$^2$, 10×10cm$^2$, 15×15cm$^2$, 20×20cm$^2$, 25×25cm$^2$) beam profiles were simulated to produce PDD curves at 100 cm SSD and lateral dose profile in air at 95 cm SSD for the open field. Again, enough particles were simulated to reduce uncertainty below 1% from superficial depth to practical range of the electrons with the same phantom volume/voxel size. The absolute output of the machine with an open field at depth of max dose was measured with film (1.01±0.02 Gy/pulse) and included in the TPS. The simulated open field profiles and the PDD's for each applicator were uploaded within Eclipse under "Model Configuration" and as text files in the appropriate format and can be accessed in <https://github.com/optmed/eFLASHBeamModeltoTPS>. The absolute dose was set to 1 Gy/MU based on the measured dose from the open field dose measurement but may be changed based on the day-to-day output measurement of the LINAC. The inputs resulted in a model that minimized error between the Eclipse beam model with that produced from GAMOS and film measurement, creating its own energy spectrum and lateral/depth dose profiles as shown in figure 4. The profiles were compared with film measurements at discrete depths along the central axis and with a film suspended at 95 cm SSD (edges taped to vertical slabs of low Z cardboard material) to measure the in-air profile or outputs of the GAMOS configured model.

A forward dose calculation was conducted for a mouse whole brain treatment and an eFLASH-RT optimized plan was produced in Eclipse for a human metastatic renal cell carcinoma patient. The dose delivered to a mouse brain tumor was determined via GAMOS MC simulation of a 1.5 cm diameter Cerrobend circular cutout (figure 2a). The dose distribution from the MC simulations were verified to agree with the film measured dose for a 5×5×10 cm$^3$ water phantom and 1×1×1 mm$^3$ voxel size at 100 cm SSD (included in supplementary figure 2A). A computed tomography (CT) scan of the mouse was imported into GAMOS with the mouse positioned to irradiate the whole brain, and lung defined as the organ at risk (figure 5). The CT scan of the human was taken with geometry intended for delivery of the treatment. Right Posterior Oblique (RPO) treatments (10MeV UHDR FLASH beam and 9MeV conventional

Clinical Treatment Planning System for eFLASH RT

beam) were created in Eclipse with a 10×10cm² applicator and a patient specific cutout to conform to the tumor shape on the right rib cage. The prescribed treatment was 16 Gy single fraction, and the dose distribution from the forward calculated plan is included in figure 6. (NOTE: The human plans were created as demonstrations and will not be delivered to the anonymized patient.)

## 3. Results

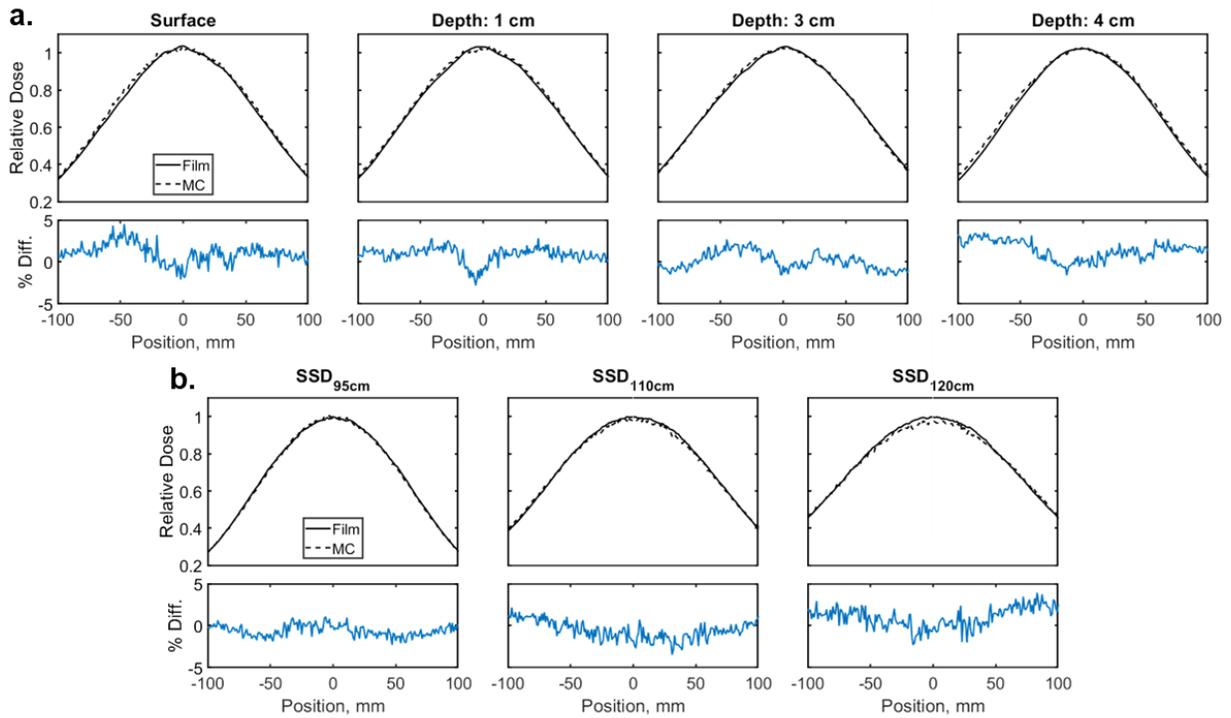

**Figure 3.** Lateral Profiles comparing GAMOS MC simulations and Film measured dose for **a.** different depths at 100 SSD and **b.** different SSD at the surface of a solid water phantom.

Varying the standard deviation $\sigma_E$ of simulated electron energy spectrum (with mean energy of 10 MeV) resulted predominantly in change in the maximum range and PDD at the distal edge of the phantom. The greatest agreement to film was found with $\sigma_E$=0.5MeV (within 5%). By varying the cone angle there was little change in the lateral profile for $\theta_{cone} > 4.63°$ because the primary collimator obstructed the edges of the beam. So, the functional relationship between $\sigma$ and $\theta_{cone}$ were determined with the simulated cone angles below 4.63°. The average cone angle that would produce the same $\sigma$ from the film (determined for each depth and lateral profile) was 3.9°±0.2°.

The comparison to film measured dose profiles suggested the GAMOS MC beam model and the Eclipse generated beam model accurately represented the electron UHDR beam produced by the modified LINAC. The lateral profiles for the open field beam at each depth for 100 cm SSD and at the surface for varying SSD (95 cm, 110 cm, and 120 cm) agreed on average within 1.5% as shown in figure 3. The lateral in-air profile from the Eclipse model and GAMOS model (figure 4a) agreed with the film measured in-air profile on average within 3% and the PDD agreed with the film measured dose at discrete depths generally withing 2% with the greatest disagreement at the tail-end of the profile. The lateral surface profile for the 1.5 cm circular field at the 100 cm SSD agreed on average within 2% and the PDD agreed within 3% (in supplementary figure 2A). The GAMOS MC simulation produced beams using a gaussian distribution for the energy spectrum centered at 10MeV and $\sigma_E$=0.5MeV and the Eclipse model's spectrum is peaked at 10MeV with a spread indicative of a gaussian distribution. However, in the Eclipse model about 5% of electrons with energy 5 MeV or less contribute to the dose delivered and is seen as a tail in the spectrum.

Clinical Treatment Planning System for eFLASH RT

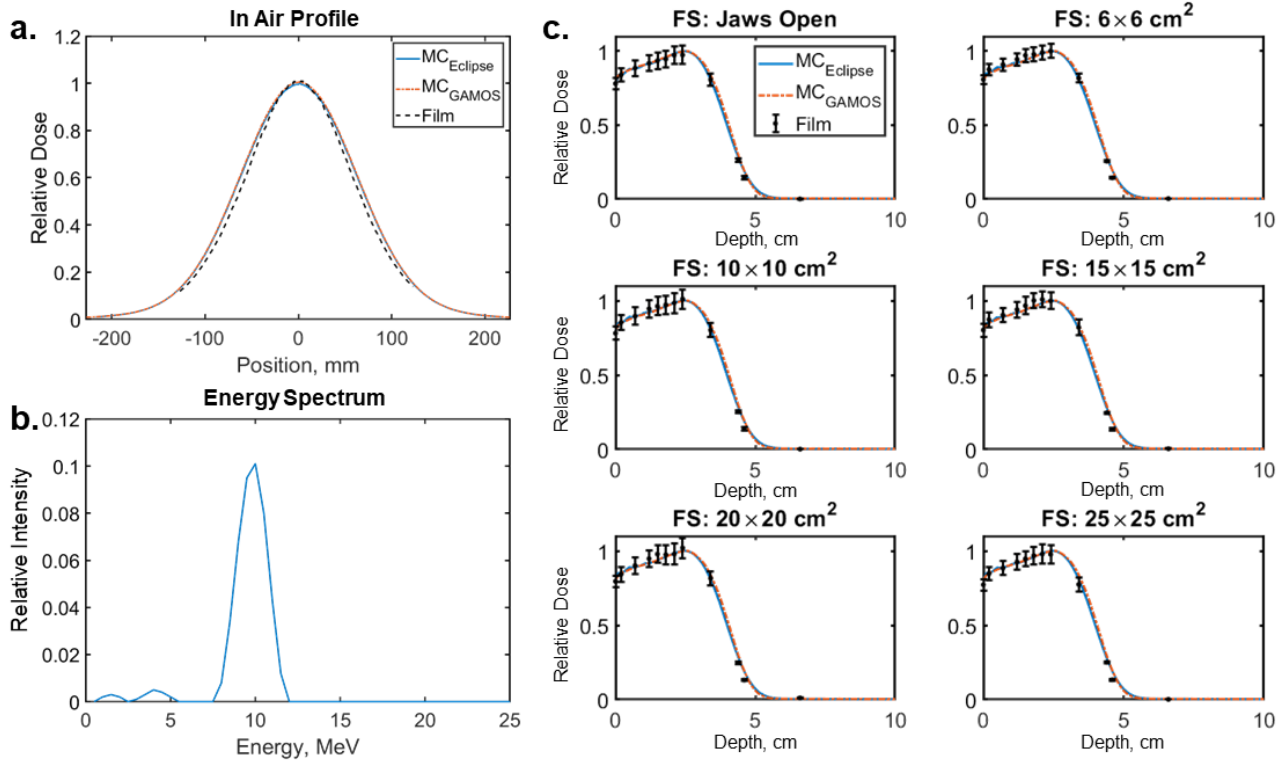

**Figure 4.** The Eclipse (Varian Medical Systems, Palo Alto, CA) calculated eFLASH beam configuration results from inputting GAMOS MC produced beam profiles, including **a.** film validated in-air profile **b.** the energy spectrum, and **c.** film validated PDD's, for all applicators and the open field.

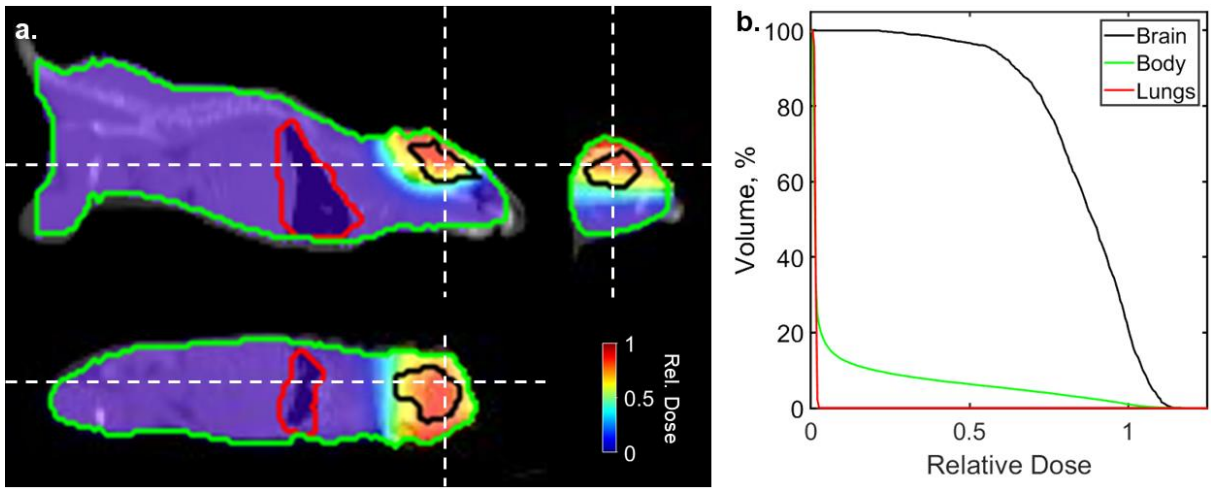

**Figure 5. a.** Dose distribution at orthogonal views and **b.** cumulative dose volume histogram for an irradiated brain of a mice with a 1.5 cm circular Cerrobend cutout from GAMOS simulation (dose distribution in water phantom is included in supplementary figure 3A). Dashed white lines in the orthogonal views indicate the slice location for the other two perspectives.

Clinical Treatment Planning System for eFLASH RT

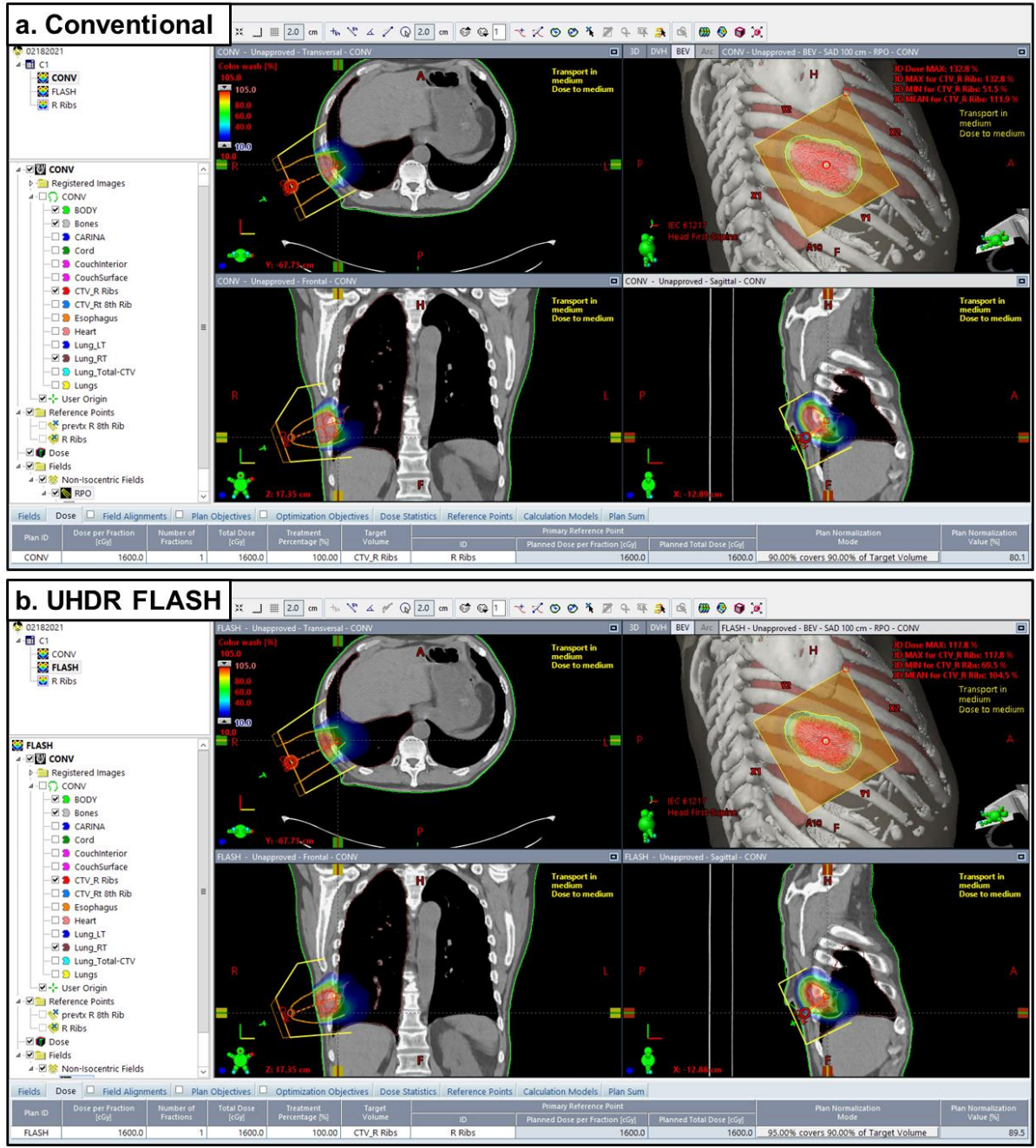

**Figure 6.** Eclipse treatment plans of a metastatic renal cell carcinoma along the right ribcage using a 10×10 cm² applicator and a cutout specifically shaped to the tumor volume. Plans include **a.** conventional 9MeV beam and **b.** UHDR FLASH 10 MeV beam. The 16Gy single treatment regimen was planned to deliver to a targeted reference point. Beams eye view (BEV) are included along with orthogonal views for each plan. (NOTE: These plans were not delivered to the patient and are included as demonstrations.)

Confirming the dose prediction from GAMOS with a water phantom for a mouse treatment (Supplementary Figure 2A) and Eclipse configured model, it was implemented to predict the dose distribution for a whole mouse brain and human renal cell carcinoma treatment as shown in figure 5 and 6, respectively. The dose volume histogram (DVH) in figure 5b suggests, while the circular field can adequately deliver dose to the entire mouse brain, the electron beam will deliver very little dose to organs-at-risk (OARs) such as the lung (or the entire body) with a steep fall off in the relative DVH. As indicated by in figure 6b, the RPO human 10MeV FLASH treatment with tumor specific cutout delivers >95% of the prescribed dose to 90% of the clinical target volume (CTV). The conventional 9MeV (nominal energy closest to the 10MeV eFLASH beam) treatment for the patient, with similar dose distribution, covers 90% of the prescribed dose to 90% of the CTV (figure 6a). The conventional plan has a greater hot spot dose (132.8% in conventional vs 117.8% in FLASH of the prescribed dose) but the eFLASH plan delivered greater dose to the right lung (11% of the lung receiving >1Gy in FLASH vs 7% of the lung receiving >1Gy in Conventional). Nonetheless, the plans suggest the eFLASH beam can adequately deliver dose to the tumor comparable to conventional electron plans.

## 4. Discussion

The agreement between measurements and calculations in GAMOS and Eclipse suggests the beam model can be used to predict the dose distribution for eFLASH-RT in preclinical and eventually clinical studies. GAMOS MC provides a free solution to treatment planning via forward dose calculations. While in figure 5 the dose distribution was used to predict the dose delivered to a mouse brain, the Eclipse TPS implementation, as indicated in figure 6 for the human renal cell carcinoma treatment, allows the user to predict and optimize dose distributions in the standard clinical environment. However, due to the nature of Eclipse TPS, a cumulative dose distribution can be predicted and optimized, without consideration of the temporal aspects of the treatment delivery[14]. To implement the FLASH effect into the treatment planning process, future work will require consideration of the dose rate distribution and its effect on biological effectiveness particularly with the delivery of multiple fields. This is conceptually straightforward as the dose rate of a specific field is proportional to the dose delivered by that field. Therefore, the dose rate distribution can be quantified on the instantaneous, field-specific averaged as well as fraction averaged levels. Prior studies have shown such dose-rate distribution can be considered in the treatment planning process[9,15,16]. The benefit here lies in utilizing already existing clinical technology regarding both the widely used Eclipse TPS and the modified LINAC in clinical geometry to deliver UHDR treatment[10].

While there are studies suggesting the FLASH effect can improve patient outcome via reduced normal tissue damage[1–4,17], implementation and improvement in the therapeutic ratio will require the delivery modalities' use in tandem with prior developed technology (i.e. TPS) in radiation therapy such as analyzing and reducing the volume of irradiated normal tissue via treatment planning[14,18]. As the radiation therapy community mobilizes to acquire and implement UHDR irradiators (e.g. LINACs) to investigate the FLASH effect, methods laid out in this study can assist in modelling their beam line, implementation into a TPS, and its potential adoption into the clinic. To promote transparency and data sharing, the source code for creating the model is included in <https://github.com/optmed/eFLASHBeamModeltoTPS>. The methods presented on determining the beam emittance parameters and energy spectrum can be utilized to model other clinically relevant beam energies (e.g. 18 MeV).

Clinical Treatment Planning System for eFLASH RT

## 5. Conclusion

A clinical TPS was configured for eFLASH-RT of a LINAC delivering UHDR at clinical treatment room geometry and predicted dose delivery to a mouse and human patient. The GAMOS MC model and its implementation in Eclipse TPS accurately represented the dose delivery as measured by Gafchromic film and the methods presented in the study can be utilized by others to accurately model their UHDR irradiator for configuration into a TPS (<https://github.com/optmed/eFLASHBeamModeltoTPS>). In future work, dose rate distribution will be implemented into the treatment planning process and further validations will conducted in large animal studies prior to the clinical translation.

**Acknowledgements**

This work was supported by the Norris Cotton Cancer Center seed funding through core grant P30 CA023108 and through seed funding from the Thayer School of Engineering, as well as support from grant R01 EB024498. Department of Medicine Scholarship Enhancement in Academic Medicine (SEAM) Awards Program from the Dartmouth Hitchcock Medical Center and Geisel School of Medicine also supported this work.

**Conflict of Interest Statement:**

Dr. Zhang has a patent 10201718 issued, and a patent 15160576 issued. Dr. Pogue reports personal fees and other from DoseOptics LLC, outside the submitted work. Dr. Bruza reports non-financial support from DoseOptics LLC, during the conduct of the study; he also reports personal fees and non-financial support from DoseOptics LLC, outside the submitted work. In addition, Dr. Bruza has a patent 62/967,302 pending, a patent 62/873,155 pending, a patent PCT/US19/14242 pending, and a patent PCT/US19/19135 pending.

Clinical Treatment Planning System for eFLASH RT

Clinical Treatment Planning System for eFLASH RT13. Rahman M, Bruza P, Lin Y, Gladstone DJ, Pogue BW, Zhang R. Producing a Beam Model of the Varian ProBeam Proton Therapy System using TOPAS Monte Carlo Toolkit. *Med Phys*. Published online October 8, 2020. doi:10.1002/mp.14532

14. Esplen N, Mendonca MS, Bazalova-Carter M. Physics and biology of ultrahigh dose-rate (FLASH) radiotherapy: a topical review. *Phys Med Biol*. 2020;65(23):23TR03. doi:10.1088/1361-6560/abaa28

15. Marlen P van, Dahele M, Folkerts M, Abel E, Slotman BJ, Verbakel WFAR. Bringing FLASH to the Clinic: Treatment Planning Considerations for Ultrahigh Dose-Rate Proton Beams. *Int J Radiat Oncol • Biol • Phys*. 2020;106(3):621-629. doi:10.1016/j.ijrobp.2019.11.011

16. Bazalova-Carter M, Qu B, Palma B, et al. Treatment planning for radiotherapy with very high-energy electron beams and comparison of VHEE and VMAT plans: Treatment planning for VHEE radiotherapy. *Med Phys*. 2015;42(5):2615-2625. doi:10.1118/1.4918923

17. Favaudon V, Fouillade C, Vozenin M-C. Radiothérapie « flash » à très haut débit de dose : un moyen d'augmenter l'indice thérapeutique par minimisation des dommages aux tissus sains ? *Cancer/Radiothérapie*. 2015;19(6-7):526-531. doi:10.1016/j.canrad.2015.04.006

18. Paganetti H. Monte Carlo simulations will change the way we treat patients with proton beams today. *Br J Radiol*. 2014;87(1040):20140293. doi:10.1259/bjr.20140293
16

**Supplementary Material A**

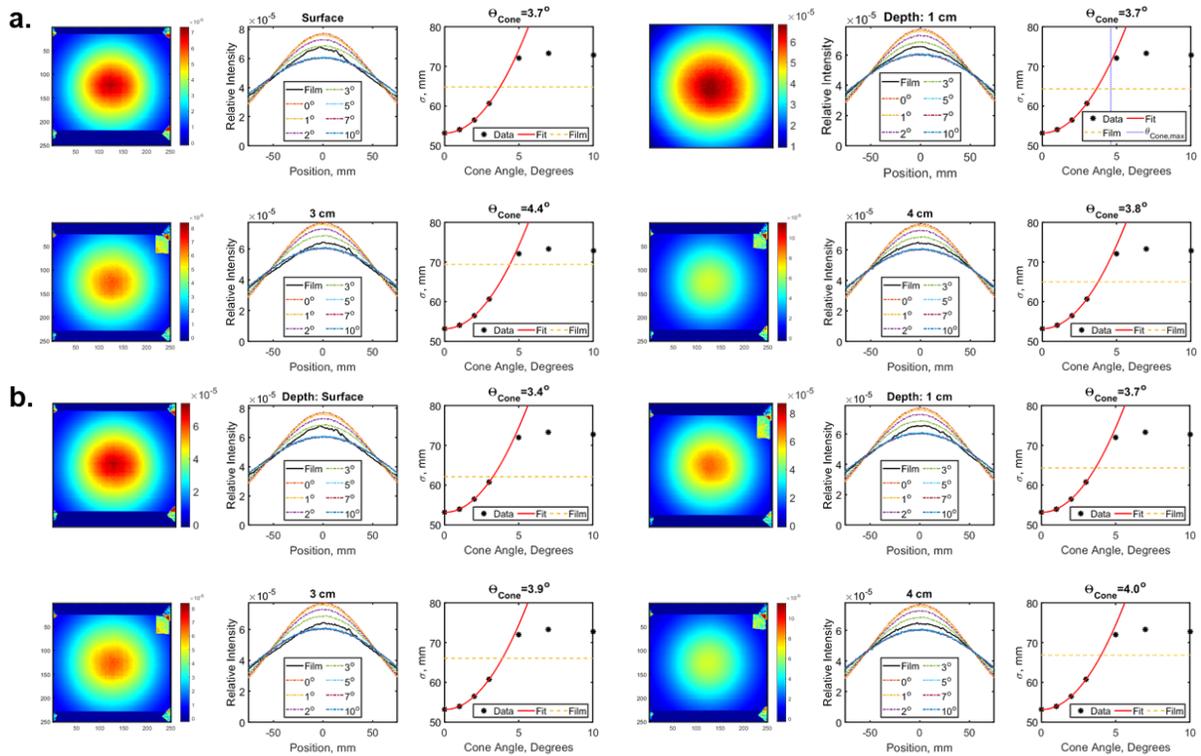

**Supplementary Figures 1A.** Determining the fitted cone angle from all **a.** cross-plane and **b.** in-plane profiles at depths of surface, 1 cm, 3cm, and 4 cm.

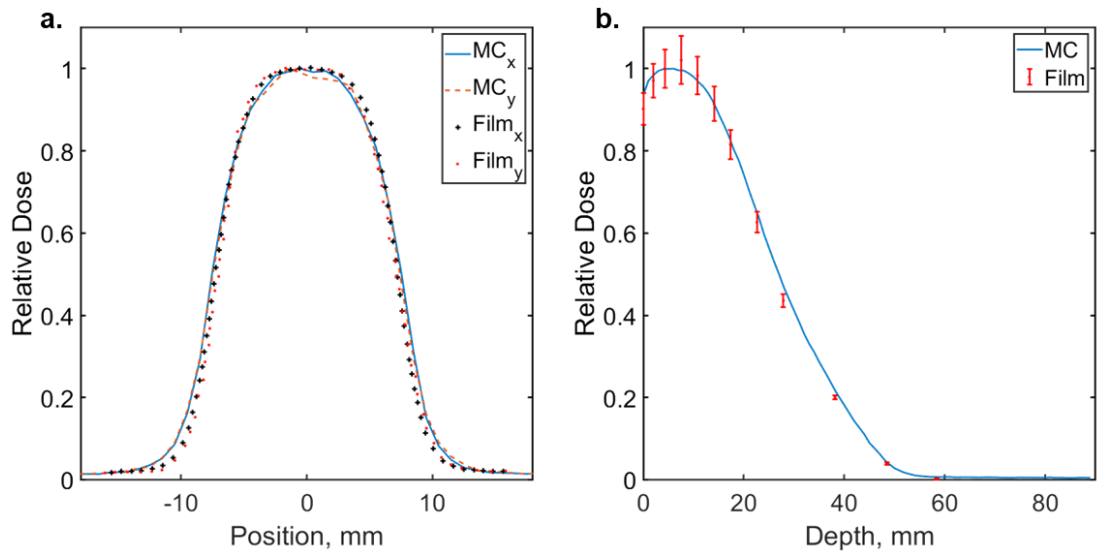

**Supplementary Figures 2A. a.** Lateral dose profile at the surface and **b.** percent depth dose curve in water phantom in comparison to film measurements (possible x-ray contamination for greater spread around penumbra and beyond).